\documentclass[conference,10pt,twocolumn]{IEEEtran}

\newif\ifdouble
\doubletrue

\usepackage[utf8x]{inputenc}
\usepackage[]{standalone}
\usepackage{tikz,pgfplots}
\usepackage{url}
\usepackage{graphicx}
\usepackage[caption=false]{subfig}
\usepackage{amsmath}
\usepackage{rotating,pdflscape}
\usepackage{booktabs}
\usepackage{multirow}
\usepackage{amsthm}
\usepackage{amssymb}
\usepackage{amsfonts}
\usepackage{wasysym}
\usepackage{algorithm}
\usepackage{algorithmic}
\usepackage[pagebackref=true,breaklinks=true,letterpaper=true,colorlinks,bookmarks=false]{hyperref}

\usepackage[numbers,sort&compress]{natbib}

\usetikzlibrary{arrows,automata,calc,shapes,shapes.symbols,positioning,fit,shadows}

\pgfplotsset{
    compat=1.12,
    every legend to name picture/.style={scale=0.7},
    every legend to name picture/.style={font=\scriptsize},
}

\definecolor{mycolor1}{rgb}{0.85,0.85,1.0}
\definecolor{mycolor2}{rgb}{1.0,0.85,0.85}
\definecolor{mycolor3}{rgb}{0.0,0.75,0.75}%
\definecolor{mycolorg}{rgb}{0.5,1.00,0.5}
\definecolor{mycolorr}{rgb}{1.0,0.5,0.5}
\definecolor{mycolorm}{rgb}{1.0,0.0,1.0}%
\definecolor{green}{rgb}{0.0,0.5,0.0}
\definecolor{lightgreen}{rgb}{0.0,1.0,0.0}

\tikzset{
    >=stealth',
    line/.style = {draw, ->},
    narline/.style = {text width=2cm, font = \small},
    operation/.style={
           rectangle,
           rounded corners,
           draw=black,
           drop shadow={shadow scale=0.95},
           fill=green!10,
           minimum width=1.4cm,
           minimum height=1.1cm,
           text width=1.65cm,
           text centered},
    flow/.style={
           diamond,
           aspect=2,
           draw=black, thick,
           fill=blue!10,
           minimum height=1em,
           text centered},
    rect/.style={
           rectangle,
           draw=black, thick,
           minimum height=1em,
           text centered},
    input/.style={
           rectangle,
           draw=black, thick,
           minimum height=1em,
           text centered}
}

\title{Large-Scale and Fine-Grained Evaluation of Popular JPEG Forgery Localization Schemes}

\author{\IEEEauthorblockN{Paweł Korus}
\IEEEauthorblockA{New York University and AGH University of Science and Technology\\ Web: \url{http://kt.agh.edu.pl/~korus/}}
}


\begin{document}

\maketitle

\IEEEpeerreviewmaketitle

\begin{abstract}
Over the years, researchers have proposed various approaches to JPEG forgery detection and localization. In most cases, experimental evaluation was limited to JPEG quality levels that are multiples of 5 or 10. Each study used a different dataset, making it difficult to directly compare the reported results. The goal of this work is to perform a unified, large-scale and fine-grained evaluation of the most popular state-of-the-art detectors. The obtained results allow to compare the detectors with respect to various criteria, and shed more light on the compression configurations where reliable tampering localization can be expected.
\end{abstract}


\section{Introduction}
\label{sec:introduction}

JPEG is the most common format for storing digital photographs and a popular subject of forensics research aiming to discover informative traces of prospective forgeries. One of the most important problems in JPEG image analysis involves distinguishing the number of compression steps that an image (or a part thereof) has undergone. Local compression inconsistencies are strong indicators of malicious forgeries. Over the years, researchers have devised multiple approaches to detect and localize such forgeries, some of which include analysis of:

\begin{itemize}
    \item pixel-domain traces aiming to reveal inconsistencies in the JPEG blocking artifacts grid;
    \item AC coefficients distribution which is distorted by subsequent re-compression;
    \item first digit distribution in AC coefficients;
    \item and most recently traces automatically learned by convolutional neural networks (CNN).
\end{itemize}

This study provides a fine-grained and large scale experimental evaluation of popular state-of-the-art tampering localization methods for JPEG images. The considered detectors (summarized in Table~\ref{tab:algorithm-summary}) represent all of the above-mentioned types of JPEG compression traces. The presented evaluation focuses on the aligned JPEG forgery scenario, i.e., where foreign content is inserted into an existing JPEG image and the resulting forgery is saved as a JPEG image with no cropping but a different quality level. In contrast to most of existing studies, which limit the quality levels to multiples of 5 or 10, the presented evaluation covers fine-grained quality level selection with differences as small as 1. Such evaluation sheds more light on the compression settings where reliable localization can be expected and on the capabilities of various detectors and their parameters.

This paper is organized as follows. Section~\ref{sec:detectors} briefly describes the considered detectors. The adopted evaluation protocol and the obtained quantitative results are presented in Section~\ref{sec:evaluation}. Section~\ref{sec:conclusions} concludes and extends the discussion by considering practical implementation issues like computational complexity and auxiliary storage requirements. Finally, Appendix~\ref{sec:mbfdf-aware-oblivious} discusses the penalty of training oblivious detectors unaware of the JPEG quality level. Example localization results are collected in Appendix~\ref{sec:localization-examples}.

\def\arraystretch{1.25}
\begin{table*}[t]
    \caption{Summary of the considered JPEG forgery detectors}
    \label{tab:algorithm-summary}
    \centering
\begin{tabular}{llllcr}
    \toprule
    \textbf{Year} & \textbf{Acronym} & \textbf{Trace} & \textbf{Authors} & \textbf{Analysis Resolution} & \textbf{Tested Implementation} \\
    \midrule
    2009 & BAG & Blocking Artifacts Grid & Li et al.~\cite{li2009passive} & $8\times{}8$~px & Image Forensics Toolbox~\cite{gihub-image-forensics} \\
    2009 & CDA & DCT Coefficients Distribution Analysis & Lin et al.~\cite{Lin2009} & $8\times{}8$~px & Image Forensics Toolbox~\cite{gihub-image-forensics} \\
    2011 & I-CDA & DCT Coefficients Distribution Analysis & Bianchi et al.~\cite{bianchi2011improved} & $8\times{}8$~px & Authors' website~\cite{cspl} \\
    2012 & BG-CDA & DCT Coefficients Distribution Analysis & Bianchi and Piva~\cite{Bianchi2012} & $8\times{}8$~px & Authors' website~\cite{cspl} \\    
    2014 & FDF-A & Mode-Based First Digit Features & Amerini et al.~\cite{Amerini2014} & $64\times{}64$~px & Image Forensics Toolbox~\cite{gihub-image-forensics} \\
    2016 & FDF-($W$/F)$^a$ & Multi-scale Mode-Based First Digit Features & Korus and Huang~\cite{Korus2016TIP} & $16\times{}16$ - $128\times{}128$~px & Own implementation~\cite{korus-github-mbfdf} \\
    2017 & CAGI & Content-aware Blocking Grid Analysis & Iakovidou et al.~\cite{Iakovidou2017} & n/a & Image Forensics Toolbox~\cite{gihub-image-forensics} \\
    2017 & CNN & Pixel-domain Convolutional Neural Network & Barni et al.~\cite{barni2017aligned} & $64\times{}64$~px & Own implementation \\
    \bottomrule
    \multicolumn{6}{l}{$^a$ $W$ denotes the analysis window size, i.e., FDF-64 represents a detector operating on $64 \times 64$~px windows}
\end{tabular}
\end{table*}
\def\arraystretch{1.0}

\section{Included Detectors}
\label{sec:detectors}

This section briefly introduces the considered forensic detectors, their parameters and training protocols (see Table~\ref{tab:algorithm-summary} for a compact summary). Whenever possible, I relied on publicly available implementations from the image forensics toolbox~\cite{gihub-image-forensics,zampoglou2016large}.

In my discussion I distinguish between \emph{oblivious} and \emph{quality-aware} detection which rely on a single trained classifier and on separate classifiers for various quality levels $Q_2$, respectively. The detectors discussed in the literature typically adopt the \emph{quality-aware} approach. I investigate the performance penalty between the two approaches in Appendix~\ref{sec:mbfdf-aware-oblivious}.

\subsection{Blocking Artifacts Grid Analysis~\cite{li2009passive}}

The algorithm proposed by Li et al.~\cite{li2009passive} exploits potential inconsistencies in the blocking artifacts grid (BAG) that stems from independent processing of image blocks by the JPEG compressor and which reflects the original blocking structure with non-overlapping $8\times8$~px blocks. 

The scheme operates by detecting the blocking grid with a second-order difference filter and extracting lines matching periodicity of 8~px. The process generates a BAG image, which may be used to spot prospective forgeries. In order to detect forgeries automatically, the authors propose to compare the the BAG image values within central $6\times{}6$~px regions of each block (where no blocking artifacts are expected) with the remaining border. Such an approach should work best for non-aligned JPEG forgeries (with obvious grid discontinuities), and for aligned forgeries with low $Q_1$ and high $Q_2$ (with weak or missing blocking artifacts in the tampered region).

The resulting tampering map is not normalized, and contains scores that are widely distributed and poorly separated between the two considered classes (single and double compression, see Fig.~\ref{fig:distributions}). In order to normalize the maps to the common range $[0,1]$, I post-processed the response by a logistic non-linearity:
\begin{equation}
    f(x) = \frac{1}{1 + e^{-\phi_1{}(x+\phi_2)}}~,
    \label{eq:logit}
\end{equation}
\noindent with $\phi_2=0$ and empirically chosen $\phi_1=0.005$.

\subsection{DCT Coefficients Distribution Analysis~\cite{Lin2009}}

AC coefficients in natural images are distributed according to a generalized Gaussian distribution. In singly compressed images, the coefficients are integers but the character of the distribution remains the same, and the probabilities smoothly decay as the coefficient magnitude grows. Double compression with a different quality level introduces characteristic periodic peaks or holes in the distribution. The detector proposed by Lin et al.~\cite{Lin2009} exploits this phenomenon for localizing individual tampered blocks. 

The algorithm computes local tampering probabilities by means of naive Bayes inference, where conditional probabilities are estimated from global mixed empirical histograms (computed for individual DCT frequencies):

\begin{equation}
    P(t_j=1|x_j^i) = \frac{P(x_j^i|t_j=1)}{P(x_j^i|t_j=1) + P(x_j^i|t_j=0)}~,
\end{equation}
\noindent where $t_j=1$ denotes the binary decision \emph{tampered} for image block $j$, and $x_j^i$ represents the observed value of an AC coefficient at frequency $i$ in block $j$. Probability estimates from individual frequencies are then combined together to yield a tampering probability map with block level resolution ($8\times{}8$~px). Due to stronger quantization, high-frequency coefficients are typically not used. By default, the implementation of the algorithm in the image forensics toolbox~\cite{gihub-image-forensics} analyzes the first~15 coefficients.

The authors also discuss a tampering detection approach, which extracts 4 features from the computed tampering probability map, and trains a machine learning classifier to predict the tampering state on the image-level.

\subsection{DCT Coefficients Distribution Analysis~\cite{bianchi2011improved}}

The detector proposed by Bianchi et al.~\cite{bianchi2011improved} is an extension of the Lin's detector~\cite{Lin2009}. The improved version recognized that the empirical distribution of the DCT coefficients is actually a mixture of two components coming from the original and the tampered regions. The authors proposed a simple approximation, which allows to estimate the expected distribution of the singly compressed image with the same quality factor. The following conditionals are considered:
\begin{equation}
\begin{split}
P(x_j^i|t_j=1) & = \tilde{h}(x_j^i)~, \\
P(x_j^i|t_j=0) & = n(x_j^i)\tilde{h}(x_j^i), ~ x_j^i \neq 0~, 
\end{split}
\end{equation}
which lead to a simple detector:
\begin{equation}
P(t_j=1) = \left( \prod_{i|x_j^i \neq 0} n_i(x_j^i) \right)^{-1},
\end{equation}
\noindent where $n(x)$ is the number of bins that map to the current bin of the histogram for a given combination of $(Q_1,Q_2)$; $\tilde{h}(x_i)$ is the expected histogram for a singly compressed image with $Q_2$ (estimated from a shifted version of the image).

Similarly to the original algorithm, the available implementation of this detector also by default uses the first 15 AC coefficients.

\subsection{DCT Coefficients Distribution Analysis~\cite{Bianchi2012}}

In a follow up study, Bianchi and Piva consider two separate cases of double JPEG compression, with aligned and non-aligned blocking grids~\cite{Bianchi2012}. Similarly to the original study~\cite{bianchi2011improved}, the algorithm considers a mixture of two distributions, but uses a more accurate probability distribution models and a more elaborate Expectation Maximization (EM) algorithm for parameter estimation. 

The available implementation of the algorithm uses the first 6 AC coefficients (consistently with the recommendation from the original research paper) and returns log likelihood ratios (LLRs). In order to convert the results to tampering probabilities, I use a logistic nonlinearity~\eqref{eq:logit} with $\phi_1=0.05$ and $\phi_2=60$. The parameters were chosen empirically on a small sub-set of test images in order to enhance separation between conditional response distributions (see Fig.~\ref{fig:distributions}).

\subsection{Mode-Based First Digit Features~\cite{Amerini2014}}

Double JPEG compression disturbs not only AC coefficient statistics, but also the statistics of their first digits, which are expected to follow the Benford's law in singly compressed images~\cite{Li2008}. This detection approach treats the frequencies of successive digits as features, and trains a classifier to distinguish between single and double compression. The frequencies are counted separately for different DCT frequencies, hence the feature set is typically referred to as mode-based first-digit features (MBFDF). For the sake of notation brevity, I use a shorter acronym FDF.

The algorithm for tampering localization proposed by Amerini et al.~\cite{Amerini2014} uses frequencies of 3 digits for 9 AC coefficients (27 features in total) extracted from $64 \times 64$~px windows. The images are analyzed in a sliding window manner. The authors considered \emph{quality-aware} detection and used support vector machine (SVM) classifiers with a multilayer perceptron (MLP) kernel. The classifiers were trained on 1920 examples (per-class) extracted from 40 training images (from the UCID dataset~\cite{UCID:2004}) and representing quality factors $Q_1,Q_2 \in \lbrace 50, 55, \ldots, 95 \rbrace$. In order to enable analysis of images with arbitrary quality levels, the detector chooses the classifier for the closest available level.

In my experiments I used a publicly available implementation from the image forensics toolbox~\cite{gihub-image-forensics} with the included pre-trained classifiers. The algorithm returns distances from the classifiers' separating hyperplane, which are converted to probabilities by a logistic nonlinearity~\eqref{eq:logit} with $\phi_1=1$ and $\phi_2=0$ (see Fig.~\ref{fig:distributions} for conditional response distributions).

\subsection{Muli-scale Mode-Based First Digit Features~\cite{Korus2016TIP}}
\label{sec:multi-scale-mbfdf}

The algorithm proposed by Korus and Huang also uses MBFDF features but adopts a multi-scale approach which involves multiple detectors operating on sliding windows of various sizes (ranging from $16 \times 16$~px to $128 \times 128$px). Candidate maps are generated with full window attribution, i.e., the score for each image block is averaged from all windows that overlap it. Such obtained candidate response maps are then fused together into a single decision map which aims to combine the benefits of small-scale and large-scale analysis. The authors considered frequencies of 9 digits for 20 modes (180 features in total) and used SVMs with radial basis (RBF) kernels trained on 10,000 examples (per-class) extracted from 1,338 images (also from the UCID dataset) representing quality levels $Q_1,Q_2 \in \lbrace 50,51,\ldots,100\rbrace$. The originally described detector adopts the \emph{quality-aware} approach. Comparison with \emph{oblivious} detection can be found in Appendix~\ref{sec:mbfdf-aware-oblivious}.

The original study considered various fusion procedures to generate the final decision map. In this paper, I use the EM fusion based on a Markov random field model. This algorithm has 4 key parameters: decision bias $\alpha$, interaction strength $\beta$, threshold drift $\delta$, and candidate map rejection threshold $\rho$. The impact of these parameters on achievable localization performance is discussed in Section~\ref{fig:param-selection}. For a detailed discussion of their operation, the readers are referred to the original study~\cite{Korus2016TIP}. In this evaluation, I used a slightly extended version of the fusion model with interactions among 8 nearest neighbors and with modulated threshold drift (see~\cite{Korus2016TIFS} for details). For the sake of lower computational complexity and limited impact on quantitative evaluation, I did not use content-adaptive interactions.

\subsection{Content-aware Blocking Grid Analysis~\cite{Iakovidou2017}}

This algorithm proposed by Iakovidou et al.~\cite{Iakovidou2017} exploits blocking artifacts grid (BAG) inconsistencies to detect and localize potential JPEG forgeries. At the time of writing, there was no public description of the algorithm (the paper was still under review). I used an open source implementation available in the image forensics toolbox~\cite{gihub-image-forensics}. 

The detector produces two response maps: a regular and an inverted one. Due to poor performance of the latter, I will present the results only for the regular one.

\subsection{Convolutional Neural Networks~\cite{barni2017aligned}}

Barni et al. discuss~\cite{barni2017aligned} 3 variants of deep-learning-based detectors with automatic feature extraction from: bitmap image patches, noise residuals, and AC coefficient histograms. They consider two network architectures based on convolutional neural networks (CNN) which are minor modifications of the well-known LeNet network~\cite{lecun1998gradient}. 

\begin{table}[t]
    \caption{Deep learning network architecture of the considered JPEG forgery detector: layer size refers to the number of filters (for convolutional layers) or neurons (fully connected).}
    \label{tab:cnn}
    \centering
\begin{tabular}{lcccc}
\toprule
\textbf{Layer} & \textbf{Kernel size} & \textbf{Stride} & \textbf{\#Size} & \textbf{Padding} \\
\midrule
Convolution & $5\times{}5$ & 1 & 30 & 2 \\
Pooling     & $2\times{}2$ & 2 & -  & - \\
Convolution & $5\times{}5$ & 1 & 30 & 2 \\
Pooling     & $2\times{}2$ & 2 & -  & - \\
Convolution & $5\times{}5$ & 1 & 30 & 2 \\
Pooling     & $2\times{}2$ & 2 & -  & - \\
Fully connected & - & - & 500  & - \\
RELU        & - & - & -  & - \\
Fully connected & - & - & 2  & - \\
Softmax & - & - & -  & - \\
\bottomrule
\end{tabular}
\end{table}

In this study, I used noise residuals as the network input\footnote{I also tried the bitmap patches but could not obtain training convergence in my experimental setup.}. The architecture of the CNN is summarized in Table~\ref{tab:cnn}. The input patches are of size $64 \times 64$~px. The residuals are computed with a popular denoising filter~\cite{Mihcak1999} on the luminance component extracted from YCbCr representation. I used 200,000 image patches for training (100,000 per class) obtained from 200 diverse 2~Mpx photographs taken by the same 4 cameras as the test images (the images were not in the test set). The quality levels were chosen randomly from $Q_1,Q_2 \in \lbrace 80,81,\ldots,100\rbrace$ with $Q_2 > Q_1$. Similarly to previous approaches, the consider primarily \emph{quality-aware} detection. Evaluation of performance gap for \emph{oblivious} detection can be found in Appendix~\ref{sec:mbfdf-aware-oblivious}.

\begin{figure*}
    \includegraphics[width=\textwidth]{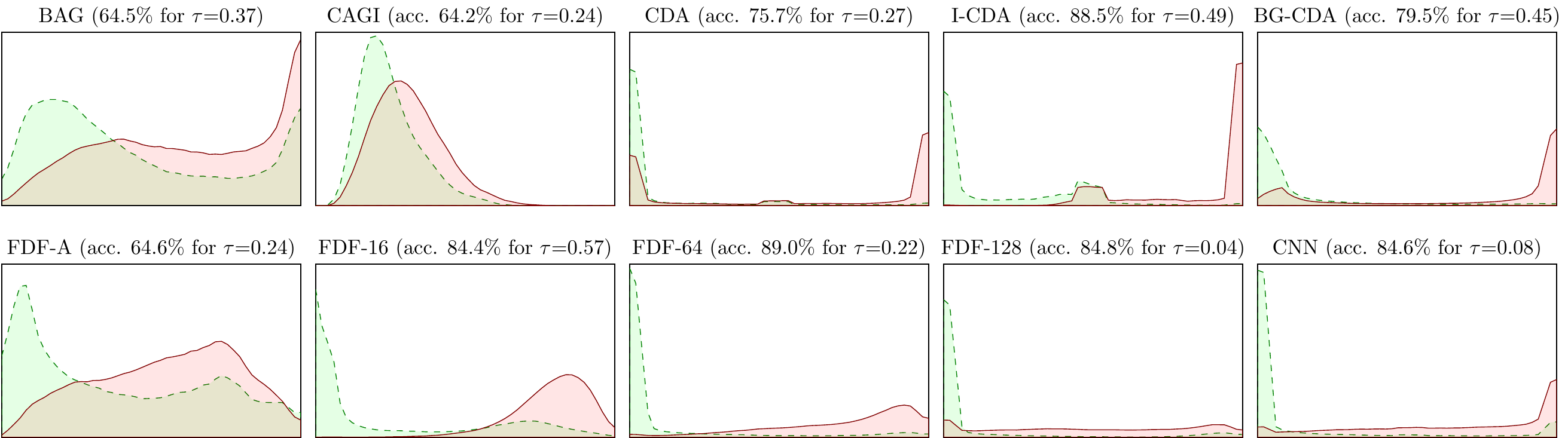}
    \caption{Conditional distributions of local detectors' responses for authentic (green) and tampered (red) areas; the distributions were obtained by randomly sampling 100 responses for 5\% of test images; the reported numbers correspond to a hypothetical classification accuracy for the given distributions and the best threshold.}
    \label{fig:distributions}
\end{figure*}

I also tried to train the network on a different dataset, e.g., on grayscale UCID images that I used for training the SVMs for the FDF detectors. However, I did not obtain satisfactory results. While the classifier appeared to be converging, it later yielded nearly random results on the test images. This suggests that this approach has rather limited generalization capabilities.

\section{Experimental Evaluation}
\label{sec:evaluation}

This section describes in detail the utilized dataset, the adopted evaluation protocol, and the obtained results.

\subsection{Dataset Preparation}

In this study, I used my recent dataset with realistic forgeries~\cite{Korus2016TIFS,Korus2016WIFS} (available for download from my website~\cite{korus-home}). The dataset contains 220 cases with various realistic forgeries including copy-move, splicing, inpainting, subtle color changes, etc. The included images are uncompressed TIFF bitmaps of size $1920\times{}1080$~px. In order to obtain test cases in the JPEG format, I automatically re-inserted tampered regions from uncompressed modified images into singly compressed original JPEG images. The resulting forgeries were then saved as JPEG with the second quality factor $Q_2 \geq Q_1$. The quality factors were chosen randomly from: $Q_1 \in \lbrace 80,81,\ldots,100\rbrace$, and $Q_2 \in \lbrace Q_1,Q_1+1,\ldots,100\rbrace$. In total, I generated 4,620 test cases with an average number of 20 cases (minimum of 9 and maximum of 36) per quality level combination $(Q_1,Q_2)$. 

\subsection{Evaluation Protocol}
\label{sec:evaluation-protocol}

All of the considered detectors yield tampering probability maps, with responses normalized to [0,1]. If an existing implementation returned different scores (e.g, LLRs), they were post-processed to obtain proper normalization (see Section~\ref{sec:detectors}). For each response map, a decision threshold was swept with 39 values uniformly distributed in [0,1]. The resulting binary maps were cleaned by removing small connected components (smaller than 4 image blocks). In case of random field-based fusion methods, no post-processing was performed. 

Each binary decision map was compared with a sub-sampled ground truth map (down to $240\times 135$ to match localization resolution of $8\times{}8$~px image blocks). The following performance metrics were collected: true positive rate, false positive rate, and $F_1$ score. The former classification rates were used to plot receiver operation characteristics (ROC) and compute area under curve (AUC) metrics for 3 maximal false positive rates - $f_p \leq \hat{f}_p : \hat{f}_p \in \lbrace 0.05, 0.1, 0.2\rbrace$, respectively, i.e.:
\begin{equation}
  \text{AUC}_{\hat{f}_p} = \frac{1}{\hat{f}_p} \int_0^{\hat{f}_p} \tilde{t}_p(f_p)~df_p
\end{equation}
\noindent where $\tilde{t}_p(f_p)$ denotes an interpolated ROC curve obtained using the piecewise cubic Hermite interpolating polynomial - a shape-preserving interpolation method. The fit was performed to match empirical $(f_p,t_p)$ samples supplemented with (0,1) and (1,0) to ensure a full curve. Division by $\hat{f}_p$ ensures score normalization to range [0,1].

\subsection{Conditional Distributions of Detectors' Responses}

Fig.~\ref{fig:distributions} shows conditional distributions of the detectors' responses obtained by randomly sampling 100 scores from 5\% of all test images. The presented histograms were mildly smoothed by a moving average filter. The reported numbers correspond to the maximal hypothetical classification accuracies estimated from the obtained distributions based on a simple threshold test (the optimal threshold $\tau$ is also given). 

It can be observed that both detectors based on blocking artifacts grid analysis (BAG and CAGI) deliver similar and rather poor separation between the scores in authentic and tampered regions. Surprisingly, the MBFDF-based detector proposed by Amerini et al. (FDF-A) also performs rather poorly and yields highly overlapping distributions. The best separation between the responses can be observed for the I-CDA and the FDF-64 detectors.

\subsection{Parameter Selection}

\begin{figure*}
    \includegraphics[width=\textwidth]{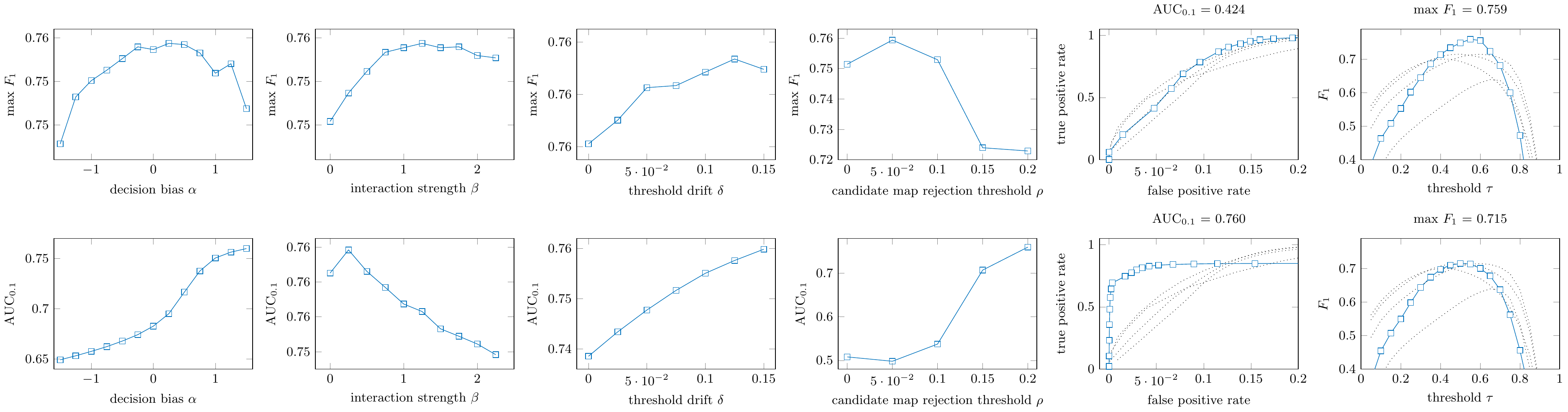}
    \caption{Impact of multi-scale fusion parameters on two localization performance metrics: $F_1$ score (top) and $\text{AUC}_{0.1}$ (bottom); Columns 1-4 show the maximal performance metric achievable for a fixed value of a single parameter; Columns 5-6 show the ROC curve and the $F_1$ score vs. decision threshold dependency; The results were obtained on 66 randomly chosen test cases and the fusion combined 4 out of 8 available candidate maps.}
    \label{fig:param-selection}
\end{figure*}

\begin{figure*}
    \centering
    \subfloat{\includegraphics[width=0.98\columnwidth]{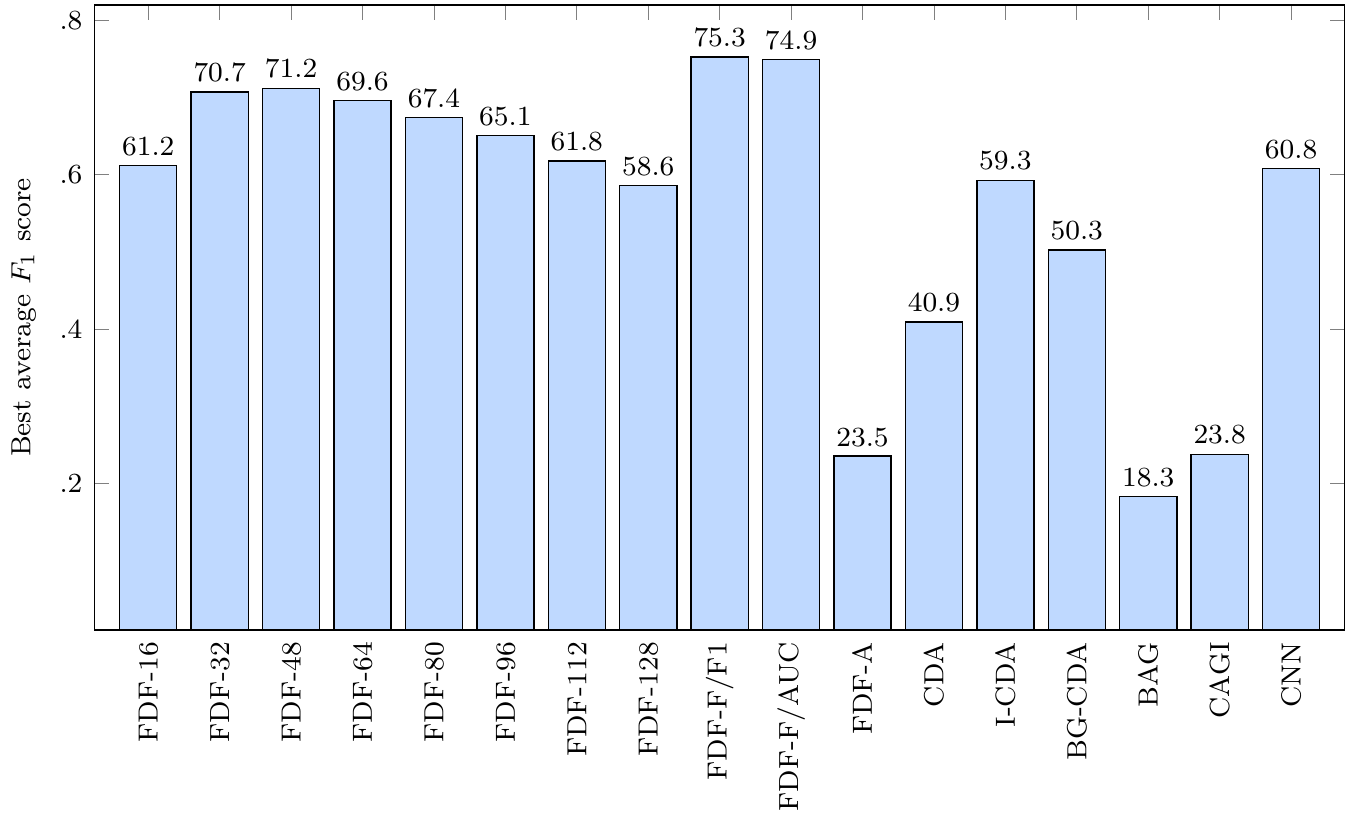}} \hspace{0.1in}
    \subfloat{\includegraphics[width=0.98\columnwidth]{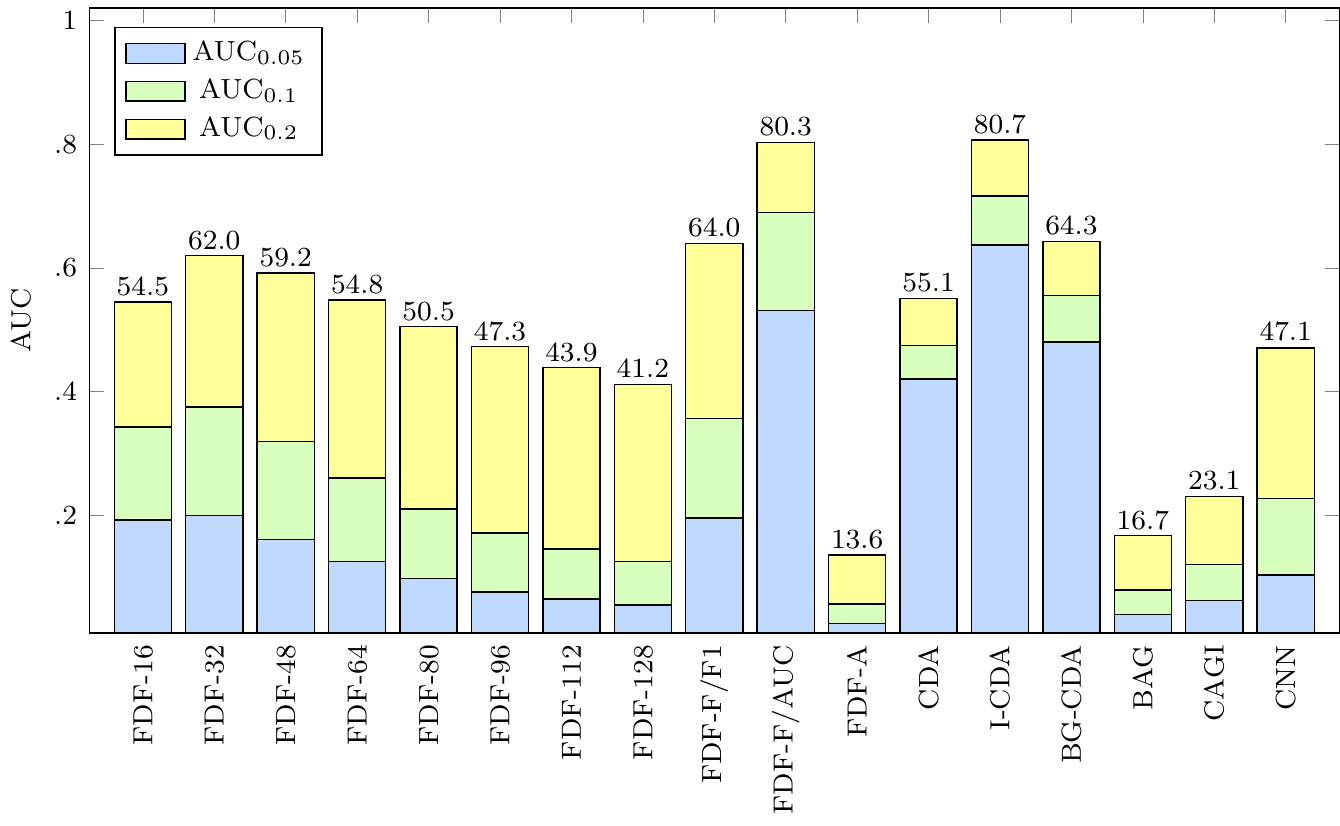}}    
    \caption{Tampering localization performance for all 4,620 test images: (left) maximal average $F_1$ scores; (right) AUC scores for $\hat{f}_p = 0.05, 0.1, 0.2$.}
    \label{fig:results-f1}
\end{figure*}

As already introduced in Section~\ref{sec:multi-scale-mbfdf}, the multi-scale fusion algorithm has four important parameters: decision bias $\alpha$, interaction strength $\beta$, threshold drift $\delta$, and candidate map rejection threshold $\rho$. In this section, I describe preliminary experiments performed to guide parameter selection.

The preliminary experiments were run on 66 randomly selected images ($\approx 1.4$\% of the whole dataset) and with only first 4 candidate maps (instead of the whole set of 8 maps). I performed grid search with the following scope:
\begin{itemize}
  \item $\alpha \in [-1.5, 1.5]$ with step 0.25,
  \item $\beta \in [0, 2.25]$ with step 0.25,
  \item $\delta \in [0, 0.15]$ with step 0.025,
  \item $\rho \in [0,0.2]$ with step 0.05,
\end{itemize}
and recorded the metrics reported in Section~\ref{sec:evaluation-protocol}. I focused on two key metrics: the maximal average $F_1$ score, and $\text{AUC}_{0.1}$. The obtained results (Fig.~\ref{fig:param-selection}) show that proper choice of the parameters allows to significantly change the behavior of the algorithm and meet different requirements. Columns 1-4 show the maximal scores achievable for a fixed choice of a single parameter. The remaining two columns show the ROC curve and a $F_1$ score vs. decision threshold $\tau$ dependency.

When optimizing the $F_1$ score, the best results were obtained for $\alpha=0.25,\beta=1.25,\delta=0.125,\rho=0.05$. The average $F_1$ score achieved the greatest value (0.759) for threshold $\tau=0.55$, whereas the best candidate map yielded 0.715 (see 6th column in Fig.~\ref{fig:param-selection}). At the same time, the ROC curve (5th column) roughly follows the best candidate maps - with slightly better performance for $f_p > 0.1$ and slightly worse for $f_p < 0.1$. 

When optimizing for best low false positive performance ($\text{AUC}_{0.1}$) the obtained results are quite different. The parameters with best performance were $\alpha=1.5,\beta=0.25,\delta=0.15,\rho=0.2$. The most important difference stems from the candidate map rejection threshold $\rho$. As can be observed in 4th column of Fig.~\ref{fig:param-selection} it has different impact for the two considered performance metrics. Greater values of the threshold mean that more unreliable candidate maps will be rejected, leading to significantly improved false positive performance and a fixed limit on the achievable true positive rate ($\approx 0.85$). For this case, the maximal obtained $F_1$ score was the same as for the best input candidate map (0.715).

Since the obtained parameter sets for the considered metrics lead to significantly different behavior, In the following evaluation I consider two variants of the detector. They will be referred to as FDF-F/F1 and FDF-F/AUC.

\subsection{Overall Localization Performance}

\begin{figure*}[!t]
    \includegraphics[width=\textwidth]{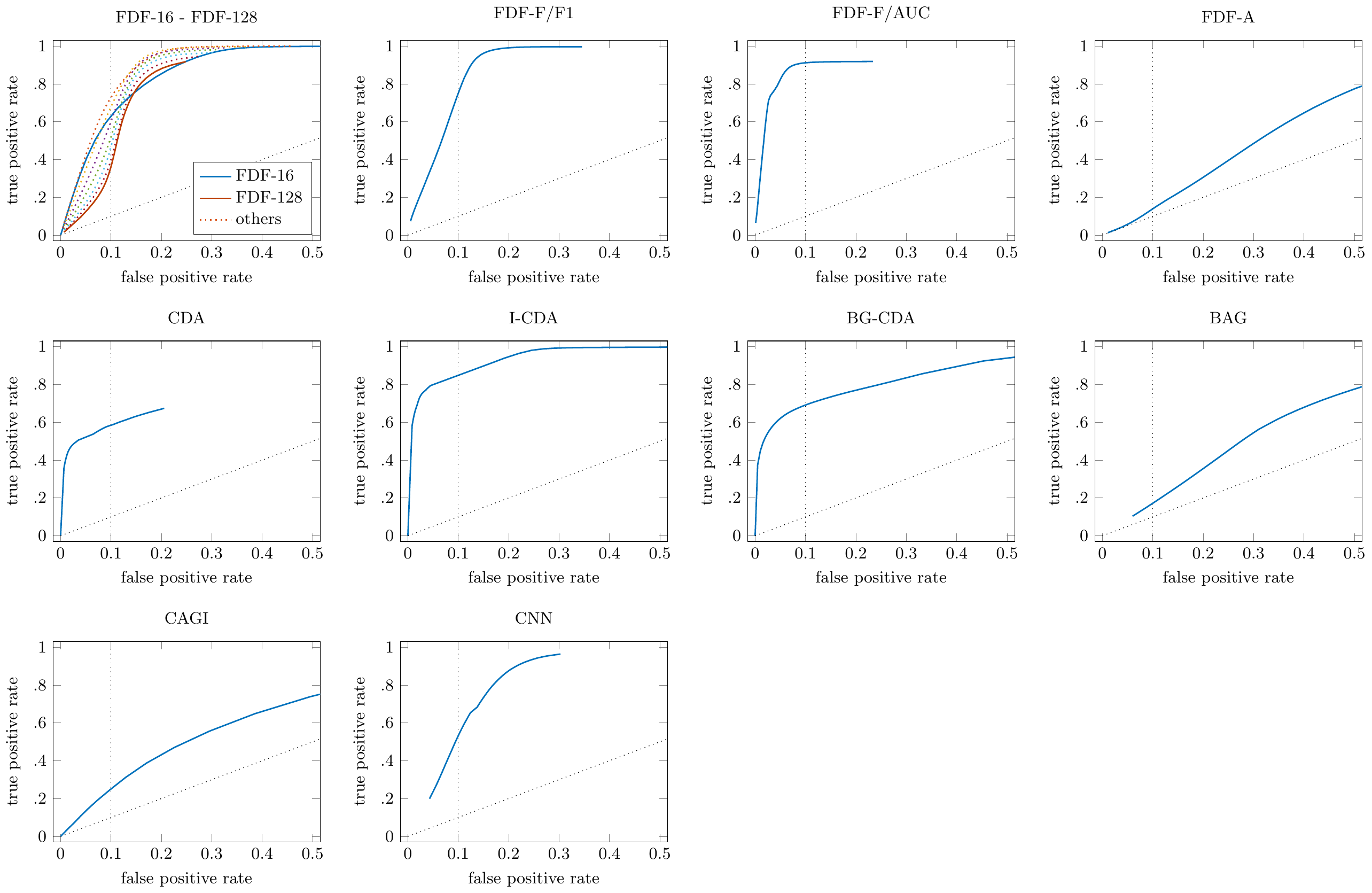}
    \caption{Receiver operation characteristics (ROC) for all 4,620 test images.}
    \label{fig:results-roc}
\end{figure*}

This section discusses overall localization performance measured for all 4,620 test images. I collected the obtained results in Fig.~\ref{fig:results-f1} and Fig.~\ref{fig:results-roc} which show the best average $F_1$ scores, the AUC scores and the receiver operation characteristics (ROC). The main observations can be summarized as follows:

\subsubsection{Blocking grid analysis}

Both algorithms based on blocking artifacts grid (BAG and CAGI) deliver very poor performance. While the newer CAGI algorithm is slightly better, the improvement is marginal and does not suffice to make it a feasible solution for dealing with aligned JPEG forgeries. Note that we expected reasonable detection results for cases with low $Q_1$ and high $Q_2$ where blocking grid discontinuity can be observed (see Section~\ref{sec:performance-quality-array} for more details).

\subsubsection{Coefficient distribution analysis}

All of the methods based on coefficient distribution analysis (CDA, I-CDA and BG-CDA) delivered good results with best performance for low false positive rates. Surprisingly, the best results were obtained for the I-CDA method which uses a simpler model of the problem than the newer BG-CDA. 

\subsubsection{First digit features}

The methods based on first digit features delivered uneven performance. The algorithm FDF-A delivered very poor results, which may indicate either an implementation problem (I used the one from the image forensics toolbox~\cite{gihub-image-forensics}), a classifier training problem, or sub-optimal scheme design (e.g, insufficient number of features). Note that I used pre-trained classifiers included in the image forensics toolbox (trained as described in the original paper~\cite{Amerini2014}) which were available for $Q_2 \in \lbrace 50, 55, \ldots, 95\rbrace$. Remaining quality levels were processed by the closest available classifier. This explanation would manifest itself as a regular grid in the results arranged by $(Q_1,Q_2)$ configurations (see Section~\ref{sec:performance-quality-array}). 

On the other hand, the set of multi-scale detectors (FDF-16 - FDF-128) delivered superior $F_1$ scores and decent ROC performance. The main problem appears to be a considerable shift towards higher false positive rates caused by unreliable response maps for some of the more challenging compression settings. As already observed in Section~\ref{fig:param-selection} on parameter selection, this problem can be successfully addressed by discarding unreliable maps - as performed by the multi-scale fusion procedure. Surprisingly, both $F_1$-driven and AUC-driven parameter configurations yielded similar $F_1$ scores - both considerably better than for any of the candidate maps. As expected, the AUC-driven configuration yielded much better results for low false positive rates - on par with the best CDA schemes.

\subsubsection{Deep learning} 

The CNN-based detector delivers competitive $F_1$ score performance, but tends to generate noisy and unreliable response maps for challenging $(Q_1,Q_2)$ configurations (see Section~\ref{sec:localization-examples} for examples). This leads to impaired low false positive rate performance. In contrast to FDF detectors whose unreliable maps resemble Gaussian noise centered around 0.5, the CNN-based detector often yields highly confident scores - visible as high-contrast noise (see the last example in Fig.~\ref{fig:examples}).

\subsection{Performance Variation with JPEG Quality Levels}
\label{sec:performance-quality-array}

\begin{figure*}[!t]
    \includegraphics[width=\textwidth]{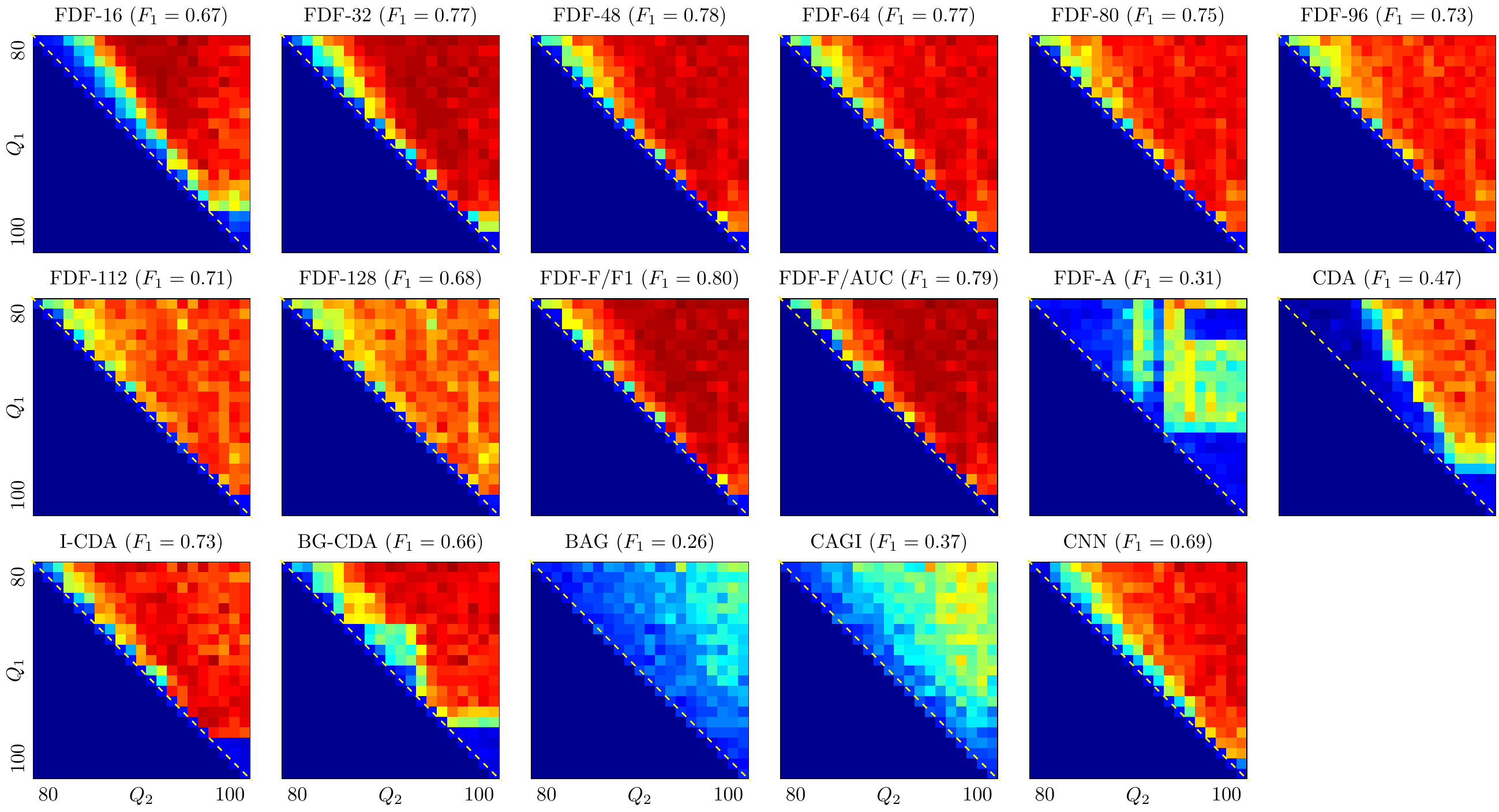}
    \caption{Localization results ($F_1$ scores) for various combinations of JPEG quality factors $Q_1$ and $Q_2$; It can be observed that detectors operating on smaller blocks deteriorate in performance when $Q_2 \approx Q_1$ and $Q_1 \geq 96$ when test images resemble singly compressed JPEGs; reported $F_1$ scores correspond to the average over all tested compression configurations.}
    \label{fig:results-array}
\end{figure*}

This section analyzes the obtained results for various compression settings. Fig.~\ref{fig:results-array} shows the average maximal $F_1$ scores (the best result for each image) for all $(Q_1,Q_2)$ configurations. The most important observation is that small-window detectors tend to have problems if $Q_1\approx Q_2$ and for $Q_1 \geq 96$. These are challenging cases which can be easily confused with uniformly single-compressed images. However, as the analysis window increases, and more reliable statistics can be collected, the reliability scope expands. This can be clearly observed for the set of multi-scale detectors, which gradually improve their scope. Windows of size $48 \times 48$~px appear to be large enough, as the scope's growth clearly saturates. Larger windows hardly improve in scope, but start to deteriorate in $F_1$ scores due to limited localization precision. Configurations with $Q_1 > 99$ remain out of reach even for the largest considered windows.

The observed results confirm our expectations towards blocking artifacts grid analysis. These algorithms visibly improve for low $Q_1$, high $Q_2$ configurations which feature weak or missing grid in the tampered areas. Even though the newer CAGI detector considerably improves upon the original BAG detector, the results are still not competitive with respect to other types of detectors.

The FDF-A detector revealed weird behavior with structured areas of somewhat better performance and zeroed scores otherwise. Surprisingly, the observed structures do not correspond to a regular grid that could have been expected from the quality levels involved in classifier training. Moreover, even for the best configurations, the performance is not competitive with respect to other detectors.

\section{Discussion and Conclusions}
\label{sec:conclusions}

The presented evaluation has shed more light on JPEG compression settings where reliable forgery localization can be expected. While it is commonly known that configurations with $Q_2 \leq Q_1$ are challenging, the remaining cases are often perceived as feasible or even easy (although cases with small difference between quality levels are known to be somewhat harder). To large extent, this simplification stems from common contraction of experimental setups to quality factors being multiples of 5 or 10. The fine-grained evaluation performed in this study assessed the performance and reliability scope of popular state-of-the-art JPEG forgery detectors, and demonstrated that many detectors deteriorate as $Q_1$ decreases, which leads to progressively wider gap between $Q_2$ and $Q_1$. Moreover, extremely high quality levels of the first compression ($Q_1 \approx 100$) are also challenging as such images have weak compression traces and could pass as uncompressed ones.

The above observations are particularly true for precise detectors which operate on individual image blocks. When larger analysis windows are used (e.g., $64 \times 64$~px), more reliable statistics can be collected, and reliability scope can improve. Overall, the reliability scope can be approximately described as follows:
\begin{equation}
\begin{split}
Q_2 & > Q_1 + \theta_1, \\
Q_1 & < 100 - \theta_2,  \\
\end{split}
\end{equation}
\noindent where $\theta_1$ is proportional to $100-Q_1$, and both $\theta_1$ and $\theta_2$ decrease with analysis window size and tend to increase for quality-oblivious detectors.

Overall, I observed the best localization results for the family of CDA detectors and for my multi-scale FDF detectors. The CDA detectors were particularly beneficial for low false positive rates. The FDF detectors delivered better $F_1$-score performance which could be further improved with multi-scale fusion. The fusion procedure could also be configured to more aggressively discard unreliable candidate maps leading to significant improvement of low false positive performance.

The key advantages of CDA detectors include the lack of detector training and low computational complexity (see Table~\ref{tab:timing} for a detailed comparison of average run-times). All classifier-based detectors incur much higher computational cost. Note that the presented results were obtained with unoptimized Matlab implementations on a common Core i7 PC (i7-4771 CPU @ 3.50GHz) and hence are susceptible to improvement by code optimization or low-level C/C++ implementation. Computational complexity could also be reduced by increasing the sliding window stride, although possibly at the cost of localization performance. The two figures reported for the CNN detector correspond to the GPU and CPU processing time respectively. The utilized graphics card was GeForce GTX 760 with 2~GB of RAM.

\begin{table}
\caption{Summary of localization performance, average processing time (2~Mpx images; Matlab implementation) and average size of classifier data (per target quality level).}
\label{tab:timing}
\centering
\begin{tabular}{lrrrr}
\toprule
\textbf{Detector} & $\mathbf{F_1}$ & $\text{\textbf{AUC}}_{\mathbf{0.1}}$ & \textbf{Time [s]} & \textbf{Storage [MB]} \\
\midrule
CDA     & .41 & .47 & 1.0 & - \\
I-CDA   & .59 & .72 & 0.5 & - \\
BG-CDA  & .50 & .55 & 4.0 & - \\
BAG     & .18 & .08 & 2.6 & - \\
CAGI    & .24 & .12 & 3.1 & - \\
FDF-A   & .23 & .06 & 72.2 & 0.3 \\
FDF-16  & .61 & .34 & 171.4 & 1.0 \\
FDF-32  & .71 & .37 & 137.2 & 2.3 \\
FDF-48  & .71 & .32 & 94.4 & 3.3 \\
FDF-64  & .70 & .26 & 72.4 & 4.1 \\
FDF-80  & .67 & .21 & 60.5 & 4.7 \\
FDF-96  & .65 & .17 & 57.7 & 5.3 \\
FDF-112 & .62 & .15 & 51.2 & 5.7 \\
FDF-128 & .59 & .13 & 49.9 & 5.9 \\
CNN     & .61 & .23 & 16.8 / 882.5$^{1}$ & 3.6 \\
\bottomrule
\\
\multicolumn{4}{l}{\footnotesize{$^1$ computation time on GPU / CPU}}
\end{tabular}
\end{table}

Classifier-based detectors also require storage space for the trained classifier parameters. This may be particularly expensive for \emph{quality-aware} detectors, where separate classifiers are trained for each JPEG quality level. The average size of classifier data, stored in Matlab's *.mat files, is collected in Table~\ref{tab:timing}. \emph{Oblivious} detectors allow to significantly reduce storage requirements, but bare considerable performance penalty (see Appendix~\ref{sec:mbfdf-aware-oblivious} for quantitative assessment).

Finally, when considering classifier-based detectors one needs to take into account their generalization capabilities. For the described multi-scale FDF-based detectors, I observed good generalization. The original classifiers were trained on gray-scale UCID images. In this study, these classifiers performed very well for color images captured with 4 different cameras (good results were also obtained in the original study for grayscale BOSS images~\cite{Korus2016TIP}). On the other hand, I was unable to obtain similar generalization for the CNN detector. A classifier trained on gray-scale UCID images yielded random results on the color test images, regardless of the adopted color conversion procedure\footnote{I also tried reading the bitmap file followed by manual YCbCr conversion and reading the luminance channel from the JPEG file followed by inverse DCT transform.}. 

\appendices

\section{Quality-aware Versus Oblivious Detectors}
\label{sec:mbfdf-aware-oblivious}

\begin{figure}[t]
    \includegraphics[width=\columnwidth]{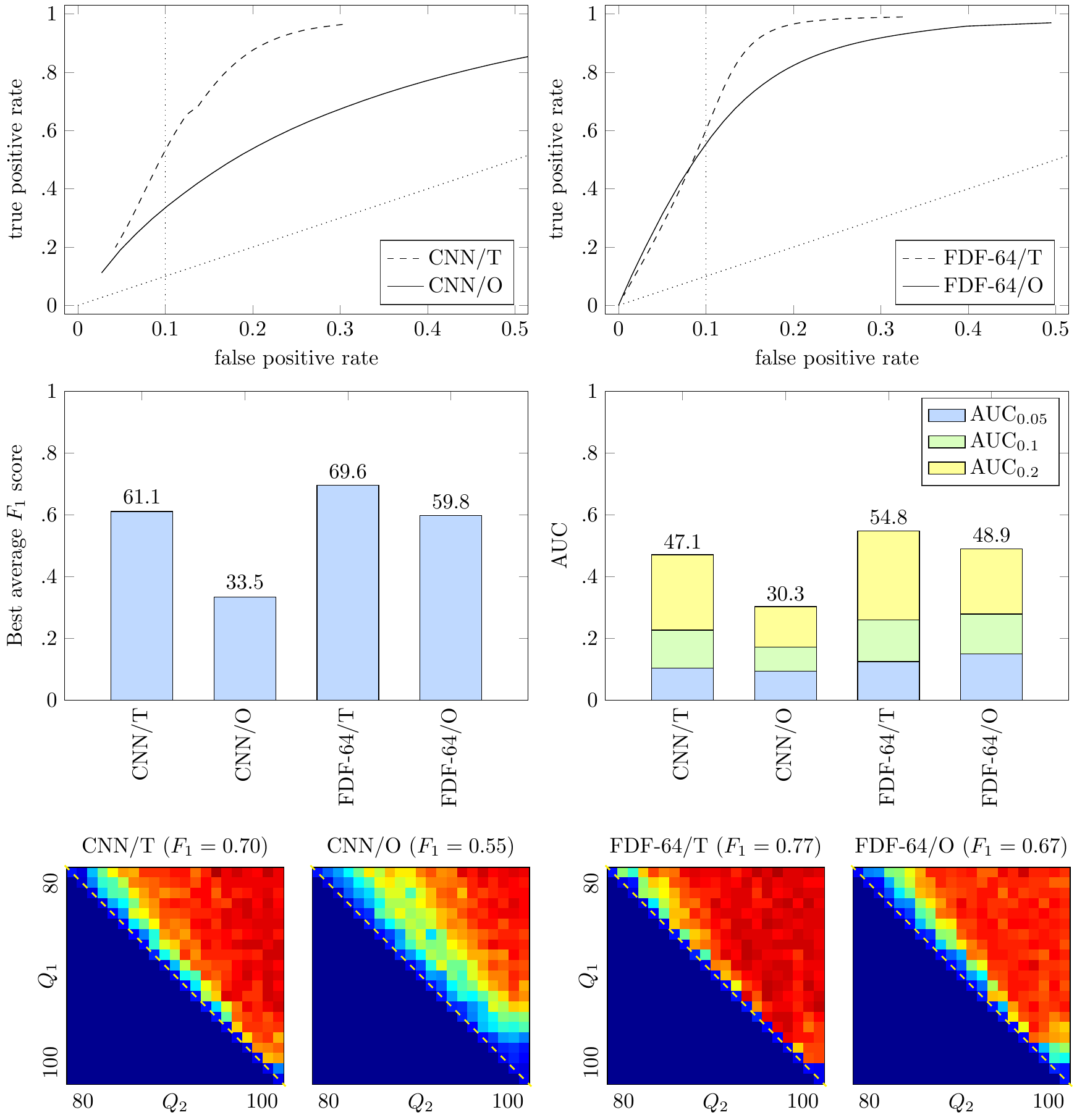}
    \caption{Comparison of tampering localization performance for quality-aware and oblivious detectors (CNN and FDF-64): (top) receiver operation characteristics; (middle) overall $F_1$ and AUC scores; (bottom) $F_1$ scores for various compression settings.}
    \label{fig:mbfdf-comparison}
\end{figure}

Quality-aware detectors incur considerable cost for storing the trained classifiers' parameters (as shown in Table~\ref{tab:timing}). One possible way of addressing this problem involves adoption of a quality-oblivious approach which uses a single, general classifier. However, this saving in storage requirements incurs a performance penalty. 

This appendix quantifies this performance penalty and presents a detailed comparison between quality-aware and oblivious versions of two detectors - the FDF and CNN detectors - both operating on $64 \times 64$~px windows. The obtained results are shown in Fig.~\ref{fig:mbfdf-comparison}. In both cases, adoption of the oblivious approach leads to considerable deterioration of all performance metrics, including $F_1$ scores, AUC scores, and reliability scopes. The scale of deterioration differs between the detectors - the FDF detector generalizes visibly better. 

\section{Localization Examples}
\label{sec:localization-examples}

\begin{landscape}
\begin{figure}[!p]
    \centering
    \includegraphics[width=9in]{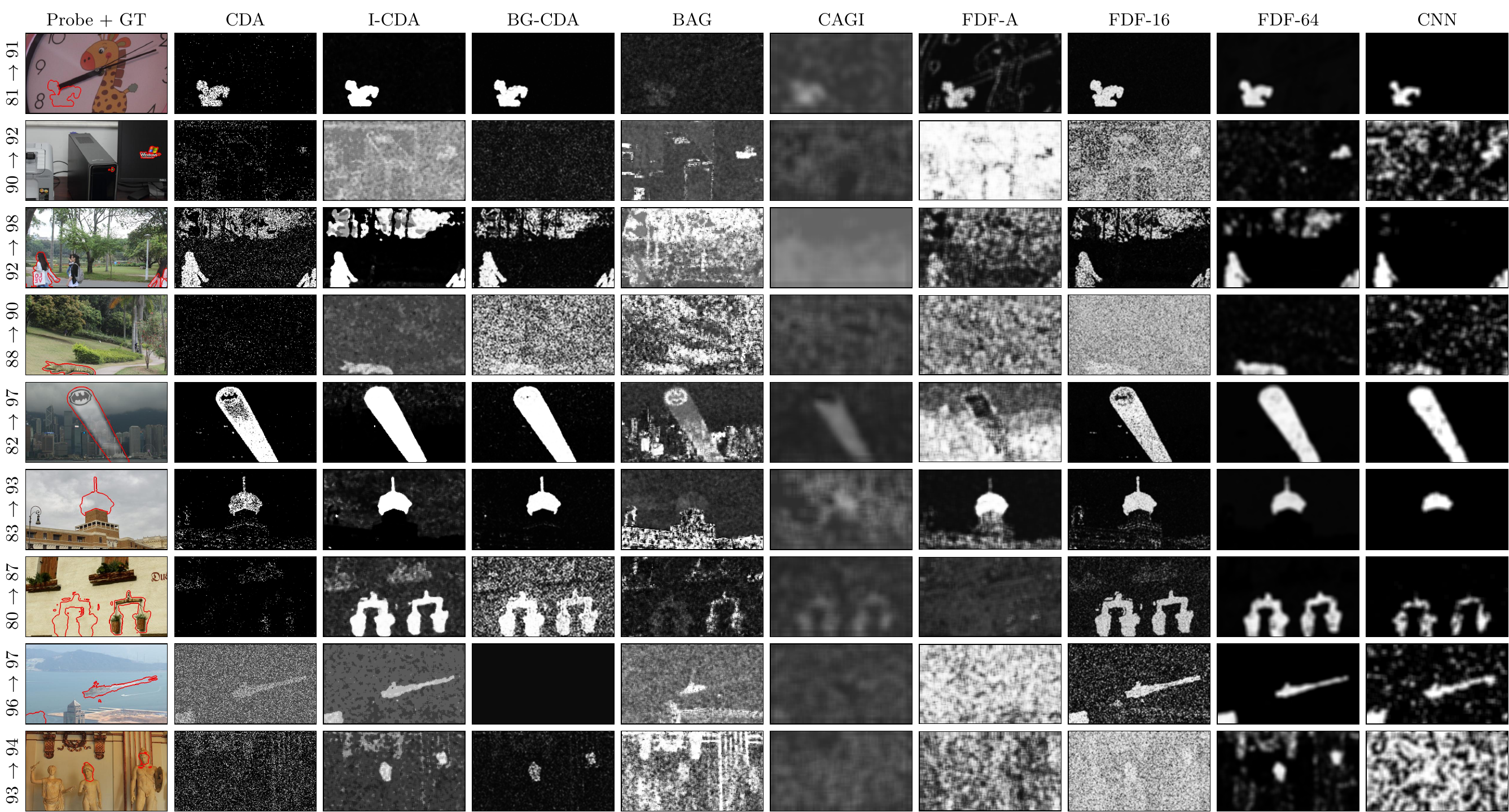}
    \caption{Tampering probability maps obtained with the considered detectors for example probe images with various compression settings (see vertical labels); ground truth maps are overlaid in red on the tampered image (1st column).}
    \label{fig:examples}
\end{figure}
\end{landscape}

Example tampering probability maps for all of the considered detectors are shown in Fig.~\ref{fig:examples}. The included examples cover several configurations of JPEG compression to demonstrate the behavior of the detectors in various conditions. It can be observed that small-window detectors, e.g., CDA or FDF-16, become less reliable when the difference between successive JPEG quality levels is small. 

Thanks to better statistics, large-window detectors, e.g., CNN or FDF-64, are more robust - both to small quality level differences and to low-information image areas. Compared to the CNN detector, whose unreliable maps tend to represent high-contrast noise, unreliable maps from the multi-scale FDF detectors often converge towards low-contrast Gaussian noise (with probabilities oscillating around 0.5).

Detectors based on blocking grid analysis perform poorly. The BAG detector typically yields completely unreliable maps with high contrast between regions of distinct content. It becomes difficult to distinguish maps with reliable localization from completely unreliable maps. The newer CAGI detectors appears to be working considerably better. While the response maps are overly smooth and attenuated, the tampered regions often form distinct shapes.

\bibliographystyle{IEEEbib}
\bibliography{general,forensics,software}

\end{document}